\documentclass[12pt,a4paper]{article}
\usepackage{graphicx}
\usepackage{fancyhdr}

\usepackage{revsymb}
\usepackage{ulem}

\usepackage{color}

\newcommand\red[1]{\textcolor{red}{#1}}

\newcommand{\rn}{{\rm n}}
\newcommand{\re}{{\rm e}}
\newcommand{\be}{\begin{equation}}
\newcommand{\ee}{\end{equation}}

\begin{document}
\def\mat#1{{\bf #1}}
\def\gmat#1{\mbox{\boldmath$#1$}}
\setlength{\headheight}{16pt}

\title{Is chemistry really founded in quantum mechanics?}

\author {Brian Sutcliffe, Service de Chimie quantique et Photophysique,
\\ Universit\'{e} Libre de Bruxelles,\\
B-1050 Bruxelles, Belgium \\and\\
 R. Guy Woolley, School of Science and Technology,\\ Nottingham Trent
 University, \\ Nottingham NG11 8NS, U.K.}

\maketitle
\section{Introduction}
\label{nintro}
Lavoisier was the first person to publish an account of how the weight relationships of reagents and 
products in chemical reactions could
serve as the basis of a systematic analytical approach to
chemistry \cite{AL:89}. Measurements of changes in weight are a characteristic feature of
the quantitative study of chemical reactions; such measurements
reveal one of the most important facts about the chemical
combination of substances, namely that it generally involves fixed
and definite proportions by weight of the reacting substances, and invariably leads 
to products consistent with the laws of conservation of mass, and of definite proportions.

A knowledge of the proportions by weight of the elements in a given
pure substance is not sufficient information to fix the chemical
identity of the substance since there may be several, or many,
compounds with the same proportions by weight of their elemental
constituents; this is true of many hydrocarbon
substances which are chemically distinct yet contain one part by
weight of hydrogen to twelve parts by weight of carbon, for
example, acetylene, benzene, vinylbenzene, cyclooctatetraene \textit{etc}.
In these cases there are distinct compounds formed by two elements
that exhibit constant chemical equivalents.

At the beginning of the nineteenth century, the chemical elements
were given a new interpretation in terms of Dalton's
atomic hypothesis that marks the beginning of microscopic chemical theory. 
Henceforth the elements were to be regarded as being composed of
microscopic building-blocks, atoms, which were
indestructible and had invariable properties, notably weight,
characteristic of the individual elements. Similarly, compounds
came to be thought of in terms of definite combinations of atoms
that we now call molecules.  All molecules of a given chemical
substance are exactly similar as regards size, mass \textit{etc}. If this
were not so it would be possible to separate the molecules of
different types by chemical processes of fractionation, whereas
Dalton himself found that successively separated fractions of a
gaseous substance were exactly similar. 

Nearly 50 years of confusion followed Dalton until the Sicilian chemist
Cannizzaro outlined \cite{SC:58} a method whereby
one could reliably determine a consistent set of weights of different
kinds of atoms from the stoichiometric data associated with a set of chemical reactions, 
and he used this to define the atomic composition
of molecules.It is important to note that the atom is the smallest unit of matter required in 
the description of chemical processes. Cannizzaro's argument was based on Avogadro's hypothesis that
equal volumes of gases at the same pressure and temperature contain equal numbers of molecules. 
From the mathematical point of view the problem is indeterminate in
the sense that one cannot exclude the possibility that the `true' atomic weights
are integer submultiples of those proposed. Cannizzaro offered a partial remedy by
observing that the probability that one has the `true' weights is increased by increasing 
the amount of data about stoichiometric relations. A complete 
account of the mathematical relations that represent stoichiometry does not  
require any assumption about the nature of atoms \cite{RGWC:95}. 

An evident limitation of stoichiometry is that it is only concerned with the changes in weight that
occur in chemical reactions; it says nothing about the changes in
other properties that accompany chemical transformations.
Equally, the original atomic theory could say nothing about the
chemical affinity of atoms, why some atoms combine and others do
not, nor give any explanation of the restriction to simple
fractions in the laws of chemical combination of atoms. In order to keep track of the growth of experimental 
results, more and more transformations of compounds into other compounds,
 some further development of the theoretical framework was needed.  In the
nineteenth century the only known forces of attraction that might
hold atoms together were the electromagnetic and gravitational
forces, but these were seen to be absolutely useless for chemistry,
and so were given up in favour of a basic structural principle.

In 1875 van 't Hoff published a famous booklet which marks the
beginning of stereochemistry \cite{JHVH1:C}. Following a suggestion
of Wislicenus, van 't Hoff proposed that molecules were
microscopic material objects in the ordinary 3-dimensional space of
our sensory experience with physicochemical properties that could be
accounted for in terms of their 3-dimensional structures. For
example, if the four valencies of the carbon atom were supposed to
be directed towards the corners of a tetrahedron, there was a
perfect correspondence between predicted and experimentally prepared
isomers, and a beautiful structural explanation for the occurrence
of optical activity. It is natural to extend this hypothesis to all molecules and 
to suppose that optically active molecules are simply distinguished from other species 
in that they possess structures that are dissymmetric. 

Here there is a clear implication 
for the dimensionality of the `molecular space'. In a two-dimensional world there 
would be two forms of the molecule CH$_2$X$_2$, whereas only one such compound is 
known. On the other hand molecules such as C-abde exist in two forms; these facts require a 
three-dimensional arrangement of the `bonds'. Evidently no
picture of the atom is required for this construction; indeed
molecular structures can be reduced to suitably labelled points
(atoms) joined by lines (bonds). Moreover van 't Hoff's
identification of ordinary physical space as the space supporting
these structures is optional; any Euclidean 3-space will do.

The development of the interpretation of chemical
experiments in terms of molecular structure was a highly original
step for chemists to take since it had nothing to do with the then
known physics based on the Newtonian ideal of the mathematical
specification of the forces responsible for the observed motions of
matter (the mechanical philosophy). It was one of the most far-reaching steps ever taken in
science. G.N. Lewis once wrote \cite{GNL1:C}
\begin{quote}
No generalization of science, even if we include those capable of
exact mathematical statement, has ever achieved a greater success
in assembling in a simple way a multitude of heterogeneous
observations than this group of ideas which we call structural
theory.
\end{quote}

Thus over a period of many years chemists developed a 
language - a system of signs and conventions for their use
- which gave them a representation of their fundamental postulate
that atoms are the building-blocks of matter; molecules are built up
using atoms like the letters of an alphabet.  A molecule in
classical chemistry is also seen as a structure, as  a
semi-rigid collection of atoms held together by chemical bonds. So
not only do we count the numbers of different kinds of atoms in a
molecule, but also we say how they are arranged with respect
to each other, and so we can draw pictures of molecules. In more abstract terms
this account has a topological quality that can be represented diagrammatically 
using `trees' or `graphs'. The laws that govern the relative dispositions of
the atoms in 3-dimensional space are the classical valency rules
which provide the syntax of chemical structural formulae.
Valency, the capacity of an atom for stable combination with
other atoms, is thus a constitutive property of the atom \cite{RGW:98}. 

The proposed atomic constitution of matter seems first to have been related to valency when
both Mendel\'{e}ev and Meyer observed, independently, in 1869
how valency was correlated with position in the periodic table 
\cite{DM91}. There was however no agreement between chemistry and physics about the nature of
atoms. In the same year as van 't Hoff inaugurated stereochemistry
with his advocacy of the tetrahedral bonding about the carbon atom, the
contributor of the entry `ATOM' in the Encyclopaedia Britannica, (generally believed to be James Clerk
Maxwell) gave strong support to Lord Kelvin's vortex model of the
atom \cite{WT1:C} because it offered an atomic model which had
permanence in magnitude, the capacity for internal motion or
vibration (which the author linked to the spectroscopy of gases), and a
sufficient amount of possible characteristics to account for the
differences between atoms of different kinds \cite{JCM1:C}.

To each pure substance there corresponds a structural molecular formula, and
conversely, to each molecular formula there corresponds a unique
pure substance. It is absolutely fundamental to the way chemists
think that there is a direct relationship between specific features
of a molecular structure and the chemical properties of the
substance to which it corresponds. Of especial importance is the
local structure in a molecule involving a few atoms coordinated to a
specified centre, for this results in the characteristic notion of a
functional group; the presence of such groups in a molecule
expresses the specific properties of the corresponding substance
(acid, base, oxidant \textit{etc}.) which however is only realized experimentally
in an appropriate reaction context. 

Each pure substance can be referred to one or several categories of 
chemical reactivity, and can be transformed into other substances which fall 
successively in other categories. The classical structural formula of a molecule summarizes 
or represents the connection between the spatial 
organization of the atoms and a given set of chemical reactions that
the corresponding substance may participate in. This set includes not
only the reactions required for its analysis and for its synthesis,
but also potential reactions that have not yet been carried
out experimentally. This leads to a fundamental distinction between
the chemical and physical properties of substances; while the latter
can be dealt with by the standard `isolated object' approach of
physics, the chemical properties of a substance only make sense in
the context of the network\footnote{Since there is no apparent limit in principle to 
the (exponential) growth in the number of new substances the chemical network is an
unbounded domain.} that describes its chemical relationships, actual and 
potential, with other substances. Of course, the chemical properties are constrained by the
physical requirements of the conservation of mass, and the competition between energy
and entropy expressed through the thermodynamic notion of `free energy'.

A notable consequence of the structural account of chemical facts
is that identical atoms can play distinguishable roles, that is the fundamental permutation
symmetry of the elemental atoms is not generally preserved in molecular structures. Thus there are
`hydroxyl hydrogens', `methyl hydrogens', `aromatic hydrogens' \textit{etc.}, and this poses
formidable conceptual problems for an `isolated object' type of description. Nevertheless, this
has been the standard approach in the physical description of molecules, as we now describe.

\section{The physical model of molecules}
\label{Intro}
Although we have seen that chemistry relies on the atom as its basic unit, there is no sufficient
theoretical account of interactions between atoms, and the conventional approach is to
invoke sub-atomic structure. Thus chemical physics and quantum chemistry rely on 
Schr\"{o}dinger's equation and an appropriate Hamiltonian for atoms and molecule 
which are taken to be composed of charged particles. 
The wave equation for the hydrogen atom with the Hamiltonian based on purely
electrostatic (Coulombic) forces between the charged particles, yields definite formulae
for the atom's energy levels expressed in terms of fundamental constants including values
for the electron and proton masses and their charges; if we substitute the
experimentally determined values (obtained from other experiments) remarkable agreement
with spectroscopic data is achieved. The quotation below comments on the `derivation' of the
wave equation for the hydrogen atom in the following terms \cite{PW:35}:
\begin{quotation}
On observing that there is a formal relation between this [Schr\"{o}dinger] wave equation
and the classical energy equation for a system of two particles of different masses and charges,
we seize on this as providing a simple, easy, and familiar way of describing the system, and
we say that the hydrogen atom consists of two particles, the electron and proton, which attract
each other according to Coulomb's inverse-square law. Actually we do not know that the
electron and proton attract each other in the same way that two macroscopic electrically
charged bodies do, inasmuch as the force between two particles in a hydrogen atom has never been
directly measured. All that we do know is that the wave equation for the hydrogen atom bears a certain formal relation to the classical dynamical equations for a system of two particles attracting each other in this way.

Having emphasized the formal nature of this correlation and of the
usual description of wave mechanical systems in terms of classical
concepts, let us now point out the extreme practical importance of
this procedure. It is found that satisfactory wave equations can
be formulated for nearly all atomic and molecular systems by
accepting the descriptions of them developed during the days of
the classical and old quantum theory and translating them into
quantum mechanical language by the methods discussed above.
Indeed, in many cases the wave mechanical expressions for values
of experimentally observable properties of systems are identical
with those given by the old quantum theory, and in other cases
only small changes are necessary. Throughout the following
chapters we shall make use of such locutions as ``a system of two
particles with inverse square attraction'' instead of ``a system
whose wave equation involves six coordinates and a function
$e^{2}/r_{12}$,'' \textit{etc}.
\end{quotation}
Since we are dealing with charged particles, a fundamental theory of atoms and
molecules must presumably be based on their electrodynamics, and so we
require electrodynamics formulated in terms of Hamiltonian dynamics, since this
is the route to Schr\"{o}dinger's equation. It is
conventional to begin with a classical description knowing
that the canonical quantization scheme due to Dirac is a
standard procedure for obtaining a quantum theory from a classical analogue that has been
cast in Hamiltonian form. It has long been recognized however that the scheme involves
analogy which may not be reliable, since the resulting quantum theory may or may not
turn out to be satisfactory. The classical theory is thus no more than a recognizable
starting point towards a quantum theory, the required endpoint.

The usual discussion in the literature of classical electrodynamics
concentrates on the Lorentz force law for the dynamics of the charges, with fields obtained
from the relevant (retarded) solutions of the Maxwell equations; much of the discussion
is concerned with aligning the theory with special relativity which is an obvious priority
in general physics. In classical electrodynamics the limiting case of point charged particles is
pathological, and a major goal of the theory is the treatment of the infinities that arise.
For example, the Coulomb energy is divergent for a classical point charge. In some sense
this means that the notion of a point particle carrying electric charge is simply inconsistent
with classical physics. We now know from  quantum mechanics that classical physics cannot
be used for lengths shorter than about the reduced Compton wavelength ($\lambdabar_C=\hbar/m_0c$) for
the particle; according to the uncertainty principle this corresponds to energies greater than
the rest-mass energy of the particle, that is above the pair production threshold. It is known
that maintaining explicit Lorentz invariance and gauge invariance provides the best route to
making sense of the divergences that plague the electrodynamics of point charged particles.

Atoms and molecules are characterized minimally by the specification of a definite number of
nuclei and electrons (molecules have `classical structures' which is a separate problem that 
will concern us below). There is no known theory of a system with a fixed finite number of 
particles interacting through the electromagnetic force that is covariant under 
Lorentz transformations, so that any general account of atoms and molecules will be 
`non-relativistic' to some degree. It is usually
accepted that the first step in transforming to a Hamiltonian description is to ensure that
Newton's law of motion for the charges with the Lorentz force, and the Maxwell
equations for the field, are recovered as Lagrangian equations of motion. There is then 
a standard calculation for the determination of the associated Hamiltonian.

It is important to note that the customary 
starting point for classical Lagrangian electrodynamics involves symbols for the
electric charges $\{e_{n}\}$ and masses $\{m_{n}\}$ of the particles which are merely
parameters that cannot be assumed to have the experimentally determined values. There is a subtle change of viewpoint here; the original equations of motion, modelled
on macroscopic classical electrodynamics, describe the electromagnetic fields associated
with prescribed sources through Maxwell's equations, while Newton's laws are
used to describe the motion of charged particles in a prescribed electromagnetic
field. The Lagrangian formalism however describes a closed system
for which $\partial L/\partial t = 0$, so that by the usual arguments the Hamiltonian $H$ is the
constant energy of the whole system.  

For comparison with experimental data the parameter $e$ is required to be the experimentally 
observed charge of a particle; a gauge invariant theory guarantees charge conservation and at 
non-relativistic energies there are no physical processes that can modify the value of $e$. 
This is true in both classical and quantum theories. The situation with the mass parameter $m$ for 
a particle is quite different since there is a charge-field interaction that leads to an
arbitrary `electromagnetic mass' additional to the `mechanical mass' $m$. It is possible
for the `electromagnetic mass' (due to self-interaction) to become
arbitrarily large and this requires $m$ to be negative so that the observed mass
$=$ mechanical mass $+$ electromagnetic mass has its observed (positive) value.
This pathology certainly occurs in the point charge limit, and is the origin
of so-called `runaway' solutions in the classical equations of
motion for the charged particles. A feature of the runaway solution is
that it has an essential singularity at $e = 0$, so there is no possibility of
constructing solutions of the interacting charge and field system that pass smoothly
into the solutions of the non-interacting system as $e\rightarrow 0$. Some of these problems
are inherited by the quantum theory resulting from canonical quantization of 
non-relativistic classical electrodynamics.

The origins of the approach employed in modern atomic and molecular theory can be found in the
model of the hydrogen atom proposed by Bohr in 1913 to account for the spectrum of hydrogen. Of course 
the model did not survive the discovery of quantum mechanics but it left a seemingly permanent imprint; the
quantum mechanics that developed from it is fundamentally spectroscopic in nature (energy levels, 
transition matrix elements, the S-matrix, response functions \textit{etc.}). In the present context it is 
perhaps worth keeping in mind an aphorism by the late Hans Primas, ``Chemistry is not 
spectroscopy'' \cite{HP:80}. Bohr's model is mainly remembered for his introduction of Planck's 
constant, $h$, and the resulting quantization of the angular momentum. Much less remarked on today is 
that Bohr made a decisive break with classical electrodynamics. In modern terms the idea is this; 
formally one fixes the gauge of the vector potential, ${\bf a}$, by the Coulomb gauge condition
\begin{equation}
\label{gauge}
\nabla \cdot {\bf a} = 0
\end{equation}
and it then follows easily that the longitudinal part of the electric field strength due to the
electrons and nuclei can be expressed entirely in terms of their coordinates and gives rise to the
familiar static Coulomb potential in the Hamiltonian $H$. `Radiation reaction' due to the transverse 
part of their electromagnetic field is discarded, and the role of the radiation field is demoted to the status 
of an `external' perturbation inducing transitions between Bohr's stationary states. Whether one can
 really separate charged particles from their own fields in a gauge-invariant fashion is another matter entirely.

The intrinsically quantum mechanical nature of the description can be seen in the characteristic 
interaction parameter, the fine structure constant\footnote{In dimensionless form 
$\alpha = e^{2}/4\pi\epsilon_{0}hc$.}, $\alpha$, which is inversely proportional to 
$h$, and so makes no sense in a classical description in which $h$ plays no role (formally $h=0$). Thus one is 
lead to the notion of an \textit{isolated} atom or molecule with a specified number of electrons and 
nuclei in free space interacting through purely Coulombic forces according to quantum mechanics.  
This framework was well-known to Schr\"{o}dinger and he too chose to
formulate his quantum theory of the hydrogen atom in terms of the singular Coulomb
potential; most mathematicians were astonished that it seemed to
work since the classical version of the model (Newtonian gravity) has pathological solutions. It 
was immediately attractive to those studying atoms and molecules, and as molecules 
had traditionally `belonged' to chemists, it was attractive to them too.

In most textbooks of physical chemistry there is, somewhere, a section
on quantum mechanics. In it there are usually examples of the use of
the Schr\"{o}dinger equation to solve a few standard
problems, at least for their bound states. These are then used to
motivate the proposition that in quantum mechanics can be found the
theoretical basis of chemistry. Although such direct assertions are
seldom to be found in textbooks of organic chemistry, quite often such
quantum mechanical constructs as `orbitals' or `potential energy
surfaces' supplement the traditional constructs of `bonds' and
`structures'.  From time to time, both `spin' and `the exclusion
principle' also get a mention. Although some of the ideas of quantum
mechanics in the Schr\"{o}dinger formulation seem to have got into
chemistry quite quickly, they were only employed in a qualitative fashion. 
It is only with the growth in widely available
computational power that the  discipline of computational quantum
chemistry became possible. This discipline claims to 
have justified Dirac's original claim that quantum mechanics could be used directly
and quantitatively to describe traditional chemical concepts, if only
the computation could be done. Let us remind ourselves of what Dirac wrote and its context. Dirac 
started by remarking that (in 1929) quantum mechanics had been nearly completed, the 
remaining problem being essentially its relationship with relativity ideas \cite{PAMD:29}. He continued:
\begin{quote}
These give rise to difficulties only when high-speed particles are involved, and are therefore of
no importance in the consideration of atomic and molecular structure and ordinary chemical reactions, in which it is, indeed, usually accurate if one neglects relativity variation of mass 
with velocity and assumes only Coulomb forces between the various electrons and atomic nuclei. The
underlying physical laws necessary for the mathematical theory of a large part of physics and the whole of chemistry are thus completely known, and the difficulty is only that the exact application of these laws leads to equations much too complicated to be soluble.
\end{quote}

The evidence for such a claim was really rather slight, probably amounting to little more than 
the work of Heitler and London on the electronic structure of the hydrogen 
molecule\footnote{Doubtless Dirac was aware of the then recent work of Born and 
Oppenheimer (1927).}, but nevertheless it has been regarded as  `received wisdom' ever since. 
Today we know that there is a well developed quantum theory for both the bound states and 
the continuum states of diatomic systems which make no contact with typical chemical ideas. 
As we shall see, an evident irony is that the claim was made in the introduction to a justly 
famous paper showing the far reaching implications of permutation symmetry in the new mechanics 
for systems of identical particles. More of that later (\S \ref{symm}). 

However that might be, it is Dirac's claim that has interested philosophers of science. We are
not philosophers but, nevertheless, we hope that we can be helpful to
them by explaining, in a not too technical way, the mathematical basis of that
claim and then putting it in a fuller context. We shall not provide
detailed references for the mathematical assertions that we make, but
we provide a few references in which the technical details
are considered and from which the original work can be identified \cite{SW:05}-\cite{SW:12b}.

In the following we shall restrict the discussion to a consideration of the bound states of the 
normal, neutral case of what may be termed the `generic' molecule. Hence we will say nothing 
about atomic and diatomic systems, and will also exclude from certain parts of the discussion,
molecules with either three or four nuclei. How calculations are or might be done is not the concern here.  
We shall leave spin properties implicit and will not consider
any relativistic effects.

The \textit{Coulomb Hamiltonian} for a system  of $N$ electrons with position variables, 
${\mat x}^{\rm e}_i$, and a set of $A$ nuclei with position variables ${\mat x}^{\rm n}_i$ 
corresponding to a given molecular formula, may be written, in the Schr\"{o}dinger representation, as
\[ \mathsf{H}({\mat x}^{\rm n}, {\mat x}^{\rm
  e})=-\frac{\hbar^2}{2m}\sum_{i=1}^{N} {\nabla}^2({\mat x}^{\rm e}_i)+
 \frac{e^2}{8\pi{\epsilon}_0}\sum_{i,j=1}^N\!\hbox{\raisebox{5pt}{${}^\prime$}}
\frac{1}{|{\mat  x}^{\rm e}_i - {\mat x}^{\rm e}_j|}
- \frac{e^2}{4\pi{\epsilon}_0}\sum_{i=1}^A
\sum_{j=1}^N \frac{Z_i}{|{\mat x}^{\rm e}_j - {\mat x}^{\rm n}_i|}\]
\begin{equation}
-\frac{\hbar^2}{2}\sum_{k=1}^{A}\frac{{\nabla}^2({\mat{x}}^{\rm n}_k)}{m_k}+
\frac{e^2}{8\pi{\epsilon}_0}\sum_{i,j=1}^A\!
\hbox{\raisebox{5pt}{${}^\prime$}}
\frac{Z_iZ_j}{|{\mat x}^{\rm n}_i - {\mat x}^{\rm n}_j|}
\label{coulham}
\end{equation}
in which the position operators are simple time-independent
multiplicative operators acting on functions of the coordinate variables (`wavefunctions'). The primes 
on the second and last summations require the diagonal ($i=j$) terms to be omitted; they represent the 
infinite self-energy of each charge referred to above. It is assumed that the charge
and mass parameters are the experimentally observed values for the particles. This Hamiltonian, together 
with its associated Schr\"{o}dinger equation
\begin{equation}
\label{CouSch}
\mathsf{H}({\mat x}^{\rm n}, {\mat x}^{\rm e})  \Psi_{n}({\mat x}^{\rm n}, {\mat x}^{\rm e}) = E_{n}\Psi_{n}({\mat x}^{\rm n}, {\mat x}^{\rm e})
\end{equation}
are taken to be the foundational equations of the quantum mechanical account of chemistry. When, 
subsequently we speak of ``the full problem'' we shall mean the theory defined by 
equations (\ref{coulham}) and (\ref{CouSch}). A short account of its properties will be 
the subject of \S \ref{ClHam}.

\section{Computational Quantum Chemistry}
\label{compQC}
In computational quantum chemistry calculations are accomplished by first clamping the 
nuclei at fixed positions and then
performing electronic structure calculations treating the nuclei as providing a classical 
potential field for the electronic motion. Thus the nuclear momentum operators must first be removed 
from (\ref{coulham}). In the second step it is proposed to reintroduce the nuclear momentum and 
position variables as quantum mechanical operators so as to accommodate the 
quantum properties of the nuclei.

With the nuclei at a particular
fixed geometry specified by the constant vectors
${\mat x}^{\rm n}_i$=${\mat a}_i$, $i=1,2,\ldots, A$, this modified Hamiltonian takes the form
\[\mathsf{H}^{\rm cn}(\mat a ,{\mat x}^{\rm e}) =
-\frac{\hbar^2}{2m}\sum_{i=1}^{N}
{\nabla}^2({\mat x}^{\rm e}_i) - \frac{e^2}{4\pi{\epsilon}_0}\sum_{i=1}^A
\sum_{j=1}^N\frac{Z_i}{|{\mat x}^{\rm e}_j - {\mat a}_i|} + \frac{e^2}
{8\pi{\epsilon}_0}\sum_{i,j=1}^N\!\hbox{\raisebox{5pt}{${}^\prime$}}
\frac{1}{|{\mat x}^{\rm e}_i - {\mat x}^{\rm e}_j|}\]
\begin{equation}
+\frac{e^2}{8\pi{\epsilon}_0}\sum_{i,j=1}^A\!
\hbox{\raisebox{5pt}{${}^\prime$}} \frac{Z_iZ_j}{|{\mat a}_i - {\mat a}_j|}.
\label{hcn}
\end{equation}

The Schr\"{o}dinger equation for the clamped-nuclei Hamiltonian\footnote{$\mathsf{H}^{\rm cn}$ is also 
commonly referred to as the `electronic' Hamiltonian.} is then
\begin{equation}
\mathsf{H}^{\rm cn}(\mat a ,{\mat x}^{\rm e})\psi^{\rm cn}_p(\mat a ,{\mat
x}^{\rm e})=E^{\rm cn}_p(\mat{a}) \psi^{\rm cn}_p(\mat a ,{\mat x}^{\rm e})
\label{cnp}
\end{equation}
in which the eigenvalues (`electronic energies') have a parametric dependence on the 
constant nuclear position vectors $\mat{a}$=$\{\mat {a}_i$\}. It is customary to omit 
the last term (the nuclear repulsion energy) from equation (\ref{hcn})
since it is merely an additive constant and so does not
affect the form of the electronic wavefunctions. Its inclusion modifies the
spectrum of the clamped-nuclei Hamiltonian only trivially by
changing the origin of the energy. 

The electronic energies are regarded as providing the energies of
chemical interest at a particular nuclear geometry specified by the
set of nuclear position vectors \{$\mat {a}_i$\}.  Thus if the
electronic energy has a minimum at a particular nuclear geometry, then
this is regarded as the geometry of a stable system. Computing the
electronic energies in the nuclear region around that minimum, is
regarded as describing a potential in which nuclei can vibrate, while
the geometry at the minimum determines the axes about which the system
can rotate. The possibility of there being a number of minima in the
potential so calculated, means that `reactants' and `products' can
be described in terms of these minima and hence transition states and
the like. Although the subsequent discussion will often seem to be in
terms of exact solutions, the actual outcome of quantum chemical
calculations will be solutions that approximate the exact ones.

However that may be, it can be shown that if the system described by the clamped-nuclei 
Hamiltonian is either neutral or positively charged, then it has an infinite number of bound 
states (an infinite set of square-integrable eigenfunctions) whatever the nuclear geometry. 
With appropriate charge and mass parameters it could thus certainly describe all molecules and 
even a system of two equal but oppositely charged parts. However a negatively charged system 
can have at most a finite number of bound states, so it is not clear if it could describe 
an isolated negatively charged ion. It has been possible to show that H$^-$ has just one bound 
state but not much else is known. In any case the spectrum is bounded from below so it is 
always possible to identify any minima in the energy.

It should be noted that the eigenfunctions \{$\psi^{\rm cn}({\mat{a},
\mat{x}^{\rm e}})$\} of the Hamiltonian, equation
(\ref{hcn}), are defined only up to phase factors of the form
\[\exp[iw(\mat{a})] \]
where $w$ is any single-valued real function of the \{$\mat{a}_i$\}
and can be different for different electronic states. It is only by making 
suitable phase choices that the electronic wavefunction is made a
continuous function of the formal nuclear variables, $\mat{a}$.
According to quantum mechanics the
eigenfunctions of (\ref{coulham}) are single-valued functions by
construction with arbitrary phases (rays). Thus care must be taken to make phase-independent comparisons when attempting to tie the clamped-nuclei Hamiltonian to the full Coulomb one.

From a mathematical point of view the eigenfunctions of the
clamped-nuclei Hamiltonian at a particular geometry, constitute a
vector bundle defined on the base space of which $\mat{a}$ is a
point. To achieve properly defined eigenfunctions for any nuclear
geometry, a suitable base space must be defined. This can be done in
the present context by defining its origin by choosing one of the
$\mat{a}_i$ to be zero and by choosing the rest of the \{$\mat{a}_i$\} to
constitute a Cartesian space. This ensures that base space is
translationally invariant and that the required vector bundle is a
trivial one. The resulting eigenfunctions are then, in terms of the
nuclear variables, single-valued and well-behaved. This space is
larger than is required to describe a geometry, for it also allows
description of all orientations and inversions of any geometrical
figure. In this space therefore, each eigenvalue of the electronic
Hamiltonian  remains the same on a spherical shell  described
by the rotation-inversion of a defined geometry. From this point of
view, the electronic Hamiltonian obviously has a completely continuous
spectrum, bounded from below.

In computational quantum chemistry practice no calculations are
made on any configuration of the nuclei which differs simply by a
rotation-inversion from a configuration at which a calculation has
actually been made.  So in practice computational quantum chemistry is
carried out on a sub-space of the full Cartesian base space which is
invariant under translations and rotation-inversions of the variables
$\mat{a}_i$.  This restriction however renders the base space
non-Cartesian. The relevant restricted base space for the nuclear
variables will be of dimension 3A-6 and to cover it, all the
internuclear distances, of which there are A(A-1)/2, are required. It
is not required to choose the inter-particle distances as internal
coordinates explicitly, and any functions of them,
$\mat{q}_i(r_{ij})$, may be used. Complete cover is  possible up to
A$=4$, but beyond that no choice of the $\mat{q}$ can cover all of the
internal motion space. Care must be taken with the general internal
coordinates for not only do they describe a restricted space, but two
distinct geometries might be described by a single choice of internal
coordinates.

It is at this level that calculations in computational quantum chemistry are made over a 
range of values, 1,2,3...s... of the \{$\mat{a_s}$\} to yield a sequence of energies, $E^{\rm
  cn}_p(\mat{a_s})$.  These values, for fixed $p$, are then fitted to construct a
`surface', $V(\mat{q})$; if, for a given geometry, a value $\mat{a_r}$ that was not
in the original list, is mapped on to a value $\mat{q_p}$ then the
calculated electronic energy, $E^{\rm  cn}_p(\mat{a_r})$ should be the
same as $V(\mat{q_p})$. The function $V(\mat{q})$ is often called a
`potential energy surface' (PES) and is treated as a basis for
considering nuclear motion associated with the electronic state labelled by the index $p$.

Any consideration of nuclear
motion must involve extending the restricted space to the full space
because the nuclear kinetic energy operators are expressed in the
full space of 3A dimensions. This can be done by using a Cartesian
variable $\mat{R}$ to describe the translation of the $\mat{a}$
and a set of three Eulerian angles \{$\gmat{\phi}_m$\} to describe their
rotation-inversion. This provides the 6 extra variables required to
fill the space. In standard electronic structure calculations the
translational motion is fixed, as explained above, by fixing the value
of one $\mat{a}_i$, and so a choice of $\mat{R}$ is inherent in such
calculations. However some other choice might prove more suitable in
considering the full problem. The choice of a particular set of
Eulerian angles, however it is made, must involve a non-linear
transformation from the Cartesian space. Thus the relationship of the
space described using the \{$\gmat{\phi}_{m}$\}, to the Cartesian base space,
can be an invertible one only where the Jacobian for the
transformation does not vanish.

For example in dealing with a
triatomic it is possible to require that the three nuclei define a
triangle and hence a plane. The Eulerian angles are defined to achieve
that end and the internal coordinates may be chosen as two sides ($r_{1},r_{2}$) and the
included angle ($\theta$) of the triangle. The part of the Jacobian that arises
from the internal coordinate choice is $r_1^2 r_2^2 \sin{\theta}$
which vanishes when any of its elements become zero. So when, say,
$\theta$ is $0$ or $\pi$ the transformation ceases to be
defined. This failure shows up by terms in the Hamiltonian expressed
in the chosen coordinates becoming singular. These singuarities will,
among other things, close off any part of the internal coordinate space in
which the nuclei form a linear system.

The Hamiltonian (\ref{hcn}) has an invariance group composed of the
electronic permutation group S$_N$ and of the point group, if present,
which when considered as acting in the space $\mat{a}$ simply
interchanges the positions of nuclei with equal charges. The
requirement that the electronic wavefunction satisfy the Pauli
principle can be achieved without explicit consideration of spin as an
extra variable, because it is possible to specify, given the
spin-state of the system, the irreducible representation of the
symmetric group S$_N$ that is then required to satisfy the Pauli
principle. Since the nuclei are regarded as identifiable in
computational quantum chemistry practice, the Pauli principle is not
relevant to permutations on the space $\mat{a}$.  This is somewhat at
odds with the usual practice of treating the nuclear kinetic energy
operators as quantum mechanical operators, when considering nuclear
motion.

So computational quantum chemistry can be used to describe not only
the electronic but also the nuclear motions. However the electronic
problem must be addressed before addressing the nuclear problem. From
a mathematical standpoint, the processes used are perfectly well
defined, providing that care is taken when moving from the restricted
to the full space. An account of the full problem in a restricted
region can therefore be properly be provided using the computational
quantum chemistry approach. Thus, for example, the electric dipole
moment of a molecule is defined by first computing the expected value
of electronic position operators with the clamped-nuclei electronic
wavefunction at the equilibrium geometry, and then adding to it the
classical electric dipole moment of the nuclei.

If the sequence of Hamiltonians (the electronic one and the subsequent
nuclear motion one) used in computational quantum chemistry are
regarded as quantum mechanical  then there is no doubt that the
theoretical basis of chemical ideas can be found in quantum
mechanics. It is certainly the case that the Coulomb Hamiltonian is a
quantum  mechanical object and if it can be shown that the
computational quantum chemical scheme can be accommodated to a
correct usage of that Hamiltonian, then there can be no doubt about
the theoretical basis of chemical ideas.

It is to such a consideration that we turn next.

\section{The Coulomb Hamiltonian and the `Isolated Molecule model'}
\label{ClHam}
In this section we try to show what features exact solutions of the Schr\"{o}dinger equation for the
Coulomb Hamiltonian for a molecular system must have even though we
don't actually have any explicit solutions at our disposal. We also try
to place the solutions obtained in computational quantum chemistry
practice in a mathematically proper relationship to the exact solutions.

Since the nearest we have to exact solutions of a molecular problem
are those for the hydrogen atom, we think that it may be helpful to begin 
by looking at them. But to go further is inevitably
to involve some quite sophisticated mathematical ideas; our aim is to make 
these ideas as plain as possible. We hope
that what we say may be sufficient, even if only skimmed, to make the
conclusions to which we come, both comprehensible and plausible.

\subsection{The hydrogen atom}

The first thing to notice is that the eigenvalues of the problem
depend only on the principal quantum number, usually written $n$, and
not on the angular quantum  number $l$. This was a  surprise to the
chemists who first thought about it, since they felt that the energy
should depend upon angular motion as well as radial motion. They thus
called it an \textit{accidental degeneracy}. This
$n^2$-fold degeneracy at each level might well have been anticipated
for it had been noted by Pauli in his 1926 treatment of the atom. It
arises in the present context because an operator corresponding to the
classical \textit{Runge-Lenz} vector, commutes with the Hamiltonian and its
symmetry is such that the $n^2$-fold degeneracy is expected. The point that is 
to be made here is that a full solution of the problem
recognises the full symmetry of the problem, even if we do not.

For the hydrogen atom the expected value of $r$ (in Bohr radius units
$a_0 \approx 10^{-10}$ m) for the level labelled by $n$ and $l$ is
\[
{\overline r}= \frac{1}{2}(3n^2-l(l+1)) \approx n^2,~~n~{\rm large}
\label{rav}
\]
and for $n=10^5$ this is about $0.5$ m. This is a macroscopic
dimension so it might be expected that the probability of finding the 
atom in this state is very small. To compute that probability one
might use the standard statistical mechanics approach. Here the partition
function is defined to normalise the total probability assuming that
the levels are randomly distributed in a Boltzmann manner. 

The partition function is defined in terms of a set of discrete energy
levels as
\[
Q=\sum_n e^{-E_n/k_BT}
\]
where $k_B$ is Boltzmann's constant and $T$ the thermodynamic
temperature, and each energy $E_n$ is counted as often as it occurs.
Working in Hartree atomic units (one Hartree atomic unit $E_h$ is about 27 eV or, 
equivalently, about 2625 kJ/mol) and writing $\beta=1/k_B T$ the general term in the 
partition function is:

\[ 2\times(n^2)\exp(-\beta/2 n^2) \rightarrow 2\times(n^2)
 ~~ \textrm{as}~~ n \rightarrow \infty.\]
The partition function thus diverges if one tries to compute the
sum.  So one cannot use it to estimate probabilities

It is clear that precisely the same sort of divergence is going to
arise for any neutral atom or, indeed, for any neutral molecule if the
nuclei are treated as being clamped. This is  because for such systems
the number of bound states is infinite and the energy levels tend to 0
as the first ionization energy is approached so that the exponential
tends to 1 but the sum does not terminate. It is usually argued that
this divergence can be ignored because the probability of any of the
higher atomic levels being occupied is negligible. (At 298 K the
second term in the sum for the hydrogen atom has a value of about
$10^{-172}$.) But the series is really 
divergent and a choice of stopping point is quite arbitrary. These troubles arise 
because the system is being treated as isolated and in an infinite space. In the physical
world no system is isolated and the space available is not
infinite. We shall discuss this further in \S \ref{tfp},

It is quite often said that bound-state eigenfunctions are
continuous, differentiable everywhere and form a complete set but that
is not always the case. 

In the hydrogen atom problem the effect of the kinetic energy operator
on the eigenfunction can be seen explicitly to produce a term proportional to $1/r$ that cancels out the potential term and thus the Hamiltonian is well-defined, even at 
the divergence point $r=0$. This must imply that
the eigenfunction behaves oddly around this point. In fact the
eigenfunctions are continuous at the origin but they are not
differentiable there, as can easily be seen by examining the explicit
functions. This might be taken as a warning that it is perilous to
neglect a kinetic energy operator in the Hamiltonian, for its
presence seems essential to overcome the singularities in the
potential. 

As for completeness, if a radial function is chosen which is of the
same kind as, but not among the bound state radial eigenfunctions,
then it is  easy to
calculate the overlap integrals between the chosen function and the
eigenfunctions. One can subsequently construct the linear combination of
eigenfunctions that maximise the eigenfunction overlap. If the set
were complete, then the value 1 would be found. In typical cases, the
value is about 0.6. This might be taken as a warning that the Coulomb
eigenfunction problem has rather special features. So a result that 
depends on an expansion assuming the eigenfunctions form a complete
set is not always safely derived. 

The hydrogen atom Hamiltonian is separable in four distinct sets of
orthogonal coordinates\red{:} spherical polar, paraboloidal, ellipsoidal and
spheroconical sets. The shapes and nodal properties of the
eigenfunctions expressed in each of these coordinate sets differ
considerably, as do the way that their quantum numbers relate to the
principal one. It is only in the spherical polar system that the
traditional orbitals appear and so the adoption of the orbital as a
chemical object might seem to be the consequence of a
simple accident of coordinate choice. But no matter which set is
chosen, the eigenvalues remain the same.

The point here is that the way the eigenfunctions of the Hamiltonian look
is entirely a matter of coordinate choice. It is the eigenvalues and
operator expectation values that are constants of the problem. It is
thus perhaps unwise to attempt physical explanations in terms of the
way eigenfunctions look.

\subsection{The full problem}
\label{tfp}

In many computational quantum chemistry papers the use of
the clamped-nuclei Hamiltonian as described in the previous section is
claimed to be related to use of the full Coulomb Hamiltonian by appeal
to the work either of Born and Oppenheimer or Born and Huang. Were
those claims supportable then this section could be a very short
one. Unfortunately the works quoted lack proper mathematical
foundation. We shall examine these claims later.

The mathematical properties of the quantum mechanical Coulomb Hamiltonian are discussed
at length  in \cite{HS:00}, and we give only a summary account.
In 1951 Kato established that the Coulomb Hamiltonian, $\mathsf{H}$, (\ref{coulham}), 
is essentially self-adjoint \cite{TK:51}. The proof involved showing that the kinetic energy
operator dominated the potential energy operator. From what has been
seen in the discussion of the hydrogen atom this observation will come
as no surprise. The property of self-adjointness, which is stronger
than Hermiticity, guarantees that the time evolution
\[\Psi(t)=\exp(-i\mathsf{H}t/\hbar)\Psi(0) \]
of a Schr\"{o}dinger wavefunction  is unitary, and so conserves probability.
This is not true for operators that are Hermitian but not
self-adjoint, and this reminds us of the importance of boundary conditions in the full
specification of a physical model. An example given by Thirring is of the radial
momentum operator $-i\hbar\partial/\partial{r}$ acting on functions
$\phi(r)$, $\phi(0)=0$ with $0\leq r < \infty$ which is not self-adjoint on the infinite half-line. Thus 
one cannot simply say `the momentum operator, $-i\hbar {\rm d}/{\rm d}x$
is self-adjoint'; one must specify the domain of the operator as well.

It is customary to assume `free' boundary conditions for the Coulomb Hamiltonian
so that the configuration space is unbounded. Then the full Galilean symmetry group of an isolated system
can be realized, and the Hamiltonian (\ref{coulham}) is recognized as the time-translation generator
in that group; the other generators are the vector operators describing space translations (the total
momentum ${\bf P}$), space rotations (the total angular momentum ${\bf J}$), and the relationship
between reference frames moving at different velocities (the `booster' ${\bf K}$). They can all be
constructed as simple sums over all the constituent particles in the system. Furthermore they can be
separated into centre-of-mass and internal contributions which are uncoupled, so that the dynamics of the
centre-of-mass can be discussed quite separately from the internal (`spectroscopic') dynamics of the
particles. We refer to this specification as the `isolated atom' ($A=1$) or `isolated molecule' ($A>1$)
model. It is evidently an ideal since in the physical world one cannot avoid confinement as 
well as interactions. The hope is that their effects are sufficiently small that they can be 
regarded as `weak perturbations'. But one has to exercise care since an unsuspecting assumption of an 
infinite configuration space can lead to pathologies, as in the H-atom example in the previous section. 

The first thing that Kato did was to make an explicit separation of the centre-of-mass
motion and the internal motions. That is essential in any search for bound
states since $\mathsf{H}$ is invariant under uniform
translations; the translation group has only continuous
irreducible representations (irreps) so that $\mathsf{H}$ must have a
completely continuous spectrum. Let us denote the translationally
invariant Hamiltonian as $\mathsf{H}'$. It can be written in terms of
$N_T-1$ Cartesian (vector) variables where $N_T=N+A$. These  can be constructed
by a non-singular linear transformation of the original Cartesian
variables. As might be expected from the discussion above, whatever
the choice of coordinates made, it has absolutely no effect on the
eigenvalues of $\mathsf{H}'$. This freedom makes it possible, whenever
it is convenient for us, to choose $A-1$ Cartesian coordinates
$\mat{t}^{\rn}_i$ constructed entirely from the original nuclear coordinates
and to retain $N$ coordinates to describe the electrons simply by
setting their origins at the centre-of-nuclear mass. These 
will be denoted ${\mat t}^{\rm e}$. With this choice made for atoms,
only the electronic coordinates survive and the centre-of-nuclear
mass becomes the mass of the single nucleus. This is why the clamped-nucleus 
hydrogen atom Hamiltonian differs from $\mathsf{H}'$ for the hydrogen atom
only by the replacement of the electronic mass with its reduced mass
equivalent.

There are various ways in which the spectrum $\sigma(\mathsf{A})$
of a self-adjoint
operator $\mathsf{A}$ may be classified. The classification most useful
in molecular physics is into
\textit{discrete} and \textit{essential} parts.
The discrete
spectrum $\sigma_{\rm d}(\mathsf{A})$ is the subset of the pure
point spectrum that consists of
isolated eigenvalues of finite multiplicity. The essential
spectrum $\sigma_{\rm ess}(\mathsf{A})$ is the complement of the
discrete spectrum
\begin{equation}
\label{esssp}
\sigma_{\rm ess}(\mathsf{A})=\sigma(\mathsf{A})\verb+\+\sigma{\rm _d}(\mathsf{A}).
\end{equation}
The discrete spectrum and the essential spectrum are, by
definition, disjoint; however, although the essential spectrum is
always closed, the discrete spectrum need not be. The essential
spectrum of the Coulomb Hamiltonian consists of the absolutely
continuous spectrum and may contain a portion of the pure point
spectrum. The operator $\mathsf{H}'$ has no singular continuous spectrum. The
essential spectrum describes scattering states of the system while
the discrete spectrum describes bound states.

The  spectrum of $\mathsf{H}'$ may (but need not) have a discrete
part, and the start of the essential part is established by means
of the so-called HVZ theorem which
demonstrates that the essential spectrum can be written as
$\sigma_{ess}(\mathsf{H}') = [\Sigma,\infty)$ where $\Sigma$ is
the energy of the lowest two-body cluster decomposition
of the $N_T-1$ particle system.

Even without recourse to detailed mathematics, it is
clear that the essential spectrum of the hydrogen atom begins at
zero energy. It is absolutely continuous and does not contain any
pure point members; it describes the scattering states of a single
electron and a nucleus. For all other atoms the first ionization
energy is such that the essential spectrum begins at somewhat
below zero energy. It contains  states describing the scattering
of an electron from a singly ionized atom, two electrons from a
doubly ionized atom and so on. These states occur at energies
below zero. This part of the spectrum is often said to describe
the bound states in the continuum but is perhaps more accurately
designated as describing resonances. At energies above zero, the
spectrum is absolutely continuous and describes the scattering of
the electrons by the nucleus. This sort of description can be
generalised to the formal Hamiltonian appropriate to any molecular
formula.

However the extent of the discrete spectrum is by no
means obvious and for a Coulomb Hamiltonian describing a given
collection of electrons and nuclei the difficult technical problem
is to find out if there is any discrete spectrum at all before the
start of the essential spectrum.

There is the following fundamental theorem due to Weyl; if one has
a trial wavefunction $\Phi$ with an expectation value
$\langle \Phi|\mathsf{H}'|\Phi\rangle$ which is below the bottom of the essential
spectrum, then $\mathsf{H}'$ has at least one discrete negative
eigenvalue. 
But it is a rather limited result; to make use of it in any
particular system the start of the essential spectrum must be
determined and a trial function found that bounds this start from
below. Both are very difficult to do. At present the most that has
been proved is that the hydrogen molecule has at least one bound
state. Ordinary chemical
experience makes it seem likely that there are some atomic
combinations that do not have any bound states but, so far, there
are no rigorous results that enable it to be said that a
particular kind of neutral system has no bound states.

An examination of tables of experimental values
of electron affinities and ionisation energies leads to the
conclusion that it is very unlikely that any diatomic molecule has
an infinite number of bound states. This observation is not
inconsistent with spectroscopic experience.  The awkward problem
in the moving nuclei case is to know whether a neutral system has
any bound states at all, although, as mentioned above, the
equivalent result in the clamped-nuclei case is known.

\subsection{The symmetries of the Coulomb Hamiltonian}
\label{symm}

It should be emphasised that the position variables in equation (\ref{coulham}) simply specify field
points, and cannot generally be identified as particle coordinates
because of the indistinguishability of sets of identical
particles. Weyl and later Mackey, both
stress that in the case of sets of identical particles, in
addition to supporting the canonical quantum conditions,  the
space on which quantum mechanical operators act must be confined
to a sub-space of the full Hilbert space of definite permutational
symmetry. This means that the effect of any operator on a function
in this sub-space must be to produce another function in the
subspace. Multiplication of  a properly symmetrised function by a
single coordinate variable produces a new function which is not in
the symmetrised sub-space. Thus only operators symmetric in all
the coordinates of identical particles can properly be deployed in
the calculation of expectation values that represent observables.
Weyl says of the two particle case \cite{HW:31}:

\begin{quote}
Physical quantities have only an objective significance if they
depend \textit{symmetrically} on the two individuals.
\end{quote}

\noindent and he then goes on to generalise this conclusion to the
symmetrical form for the quantities constructed from the variables
of  $N$ identical particles. He closes his discussion by looking
at the two electron problem. He says that although it might be
supposed that the electrons as a pair of twins could be named
``Mike'' and ``Ike''

\begin{quote}
it is impossible for either of these individuals to retain his
identity so that one of them will always be able to say ``I'm
Mike'' and the other ``I'm Ike''. Even in principle one cannot
demand an alibi of an electron! In this way the Leibnizian
principle of \textit{coincidentia indiscernibilium} holds in
quantum mechanics.
\end{quote}

\noindent This discussion holds for identical particles of any
kind that are to be described by quantum
mechanics\index{quantum mechanics} and it precludes the
specification of, for example, the expected value of a particular
coordinate chosen from a set describing many identical particles.

The Hamiltonian
$\mathsf{H}'$ is also invariant under all rotations and
rotation-reflections of the translationally invariant coordinates;
it will have eigenfunctions which provide a basis for irreducible
representations (irreps) of the orthogonal group in three
dimensions O(3). Thus the eigenfunctions are expected to be of two
kinds classified by their parity; each kind consists of
eigenfunction sets, each with degeneracy $2J+1$, according to the
irrep $J= 0, 1, 2,\ldots$ of SO(3) to which the eigenfunctions
belong. The representations of O(3) are distinct for each parity,
and so there is no group theoretical reason to expect
eigenfunctions with different parity to be degenerate.

Simultaneously the eigenfunctions will provide irreps for the
permutation group $\cal{S}$ of the system. This group comprises
the direct product of the permutation group ${\cal{S}}_N$ for the
electrons with the permutation groups ${\cal{S}}_{A_i}$ for each
set of identical nuclei $i$ comprising
$A_i$ members. The physically realisable irreps of this group are
restricted by the requirement that, when spin is properly
incorporated into the eigenfunctions, the eigenfunctions form a
basis only for the totally symmetric representation, if bosons
(spin 0, 1, 2 \textit{etc.}) or of the antisymmetric representation, if
fermions (spin 1/2, 3/2, 5/2 \textit{etc.}). Both of these representations
are one-dimensional. We shall speak of irreps of the
translationally invariant Hamiltonian which correspond to
physically realisable states as \textit{permutationally allowed}. In
general such irreps will be many dimensional and so we would
expect to have to deal with degenerate sets of eigenfunctions in
attempting to identify a molecule in the solutions to the
translationally invariant problem. Unfortunately the dimensions of the
permutationally allowed representations are very large and so it is
necessary to consider eigenvalue sets of extensive degeneracy. To look
for the singlet spin of a 22 electron system, say C$_3$H$_4$, would
involve dealing with a degenerate set of size 6$\times$10$^4$.

Our primary concern here is not with the extent of the electronic
degeneracy, but that of the nuclei and to what extent the requirements
of nuclear permutations are consistent with the occurrence of
\textit{isomers}. It has already been shown that isomers may be treated
explicitly in the clamped-nuclei approach because molecular structure
can be recognised there. Consideration of these matters in the context
of solutions of $\mathsf{H}'$ will next be considered.

\subsection{Molecular structure and isomers}
\label{msiso}
The most severe problem associated with the formal
quantum mechanical description perhaps is the fact that the Coulomb
Hamiltonian (\ref{coulham}) is one and the same for all possible
isomers associated with a given chemical formula. In the words of
P.-O. L\"{o}wdin echoing Dirac's claim (\S \ref{Intro}) \cite{POL:89}:

\begin{quote}
The Coulombic Hamiltonian $\mathsf{H}$ does not provide much
obvious information or guidance, since there is [sic] no specific
assignments of the electrons occurring in the systems to the
atomic nuclei involved - hence there are no atoms, isomers,
conformations \textit{etc}.  In particular one sees no molecular symmetry,
and one may even wonder where it comes from.  Still it is evident
that all this information must be contained somehow in the
Coulombic Hamiltonian.
\end{quote}

A natural element of classical 
molecular structure theory is to assign static dipoles to
particular molecules so as to account for the difference between `non-polar' and `polar'
molecules demonstrated by the temperature behaviour of their electric susceptibilities 
(the Langevin-Debye law). Such assignments are often made in terms of
vector sums of bond dipole moments, so that bonds are deemed to
play an important role in static 
molecular dipoles. However the Coulomb
Hamiltonian commutes with the inversion operator, so its
eigenstates must be parity eigenstates and hence must have zero
expectation values for the static electric dipole operator. But if
an eigenstate corresponds to a molecule with structure then, it
follows, that the molecule cannot have a static electric dipole
moment. The result cannot be doubted, but it has a very
paradoxical flavour. There is a quantum-mechanical account of the Langevin-Debye law, given 
many years ago by van Vleck \cite{JHVV:32, RGW:76}. Its ingredients are: 
expectation values of the \textit{square} of
the electric dipole operator, transition matrix elements of the dipole operator and the energy 
level separations of states supporting fully allowed dipole transitions, and the thermal energy
$k_{B}T$. van Vleck's calculation makes no reference to bond dipoles, nor `structures' and 
probably for that reason is now much less well known than his analogous treatment of magnetic
susceptibilities.

A similar argument leads to the conclusion
that the existence of stereo-isomers cannot be accounted for in
terms of eigensolutions of the Schr\"{o}dinger
equation for the Coulomb
Hamiltonian, for the optical rotation
angle is a pseudoscalar observable (Hund's paradox). Clearly then, an eigenstate of
$\mathsf{H}'$ does \textit{not} correspond to a classical molecule
with structure ! And if one responds that chemistry is of course concerned with
\textit{time-dependent} states, the observation invites the question: what are the
equations that \textit{determine} the time-dependent quantum states of molecules ?
Unless one simply accepts the clamped-nuclei approach to the problem, 
we have no idea.

There are more difficulties; in classical structural chemistry,
different isomers mean different geometries and the idea of a
distinct geometry is problematic for the stationary states of the
Coulomb Hamiltonian. If we write the
variables corresponding to the carbon nuclei in a generic case such 
as the hydrocarbon C$_8$H$_8$ as
$\mat{x}^n_j,~j=1,\ldots8$ and those corresponding to the protons
as $\mat{x}^n_{i+8},~i=1,\ldots8$ then a particular CH
interparticle distance is
\[x^{CH}_{ij}=|\mat{x}^n_{i+8}-\mat{x}^n_j| .\]
One might be tempted to suppose that the calculation of the
expected values of such interparticle distances with a particular
eigenfunction of $\mathsf{H}'$ would determine the geometry;
however $x^{CH}_{ij}$ is not a proper observable. As noted 
earlier, the only possible operator incorporating these distances 
is the symmetrical sum
\[\sum_{i, j=1}^8 x^{CH}_{ij} \] and all that can be inferred from
its expectation value is that, on average, all the CH
interparticle distances are the same. This is not to suppose that
this average value is the same for all the eigenfunctions of
$\mathsf{H}'$ that might be investigated in a search for isomers,
it is simply that what differences there might be, cannot support
the detailed geometrical interpretation which is characteristic of
classical chemical structure theory.

It is possible to
define an electronic charge density by integrating the squared
modulus of the total wave function over all but one of the
electronic space coordinates and all of the electronic spin
coordinates and all of the nuclear variables. This process would
yield an electronic charge density function corresponding to
expected values of the nuclear variables. This would seem to be
the closest that one might get to a  clamped-nuclei result. The
density so  calculated here would reflect
precisely the nuclear permutational symmetry alluded to above and
so knowledge of the charge density would not help identify a
 molecular structure or pattern of
bonding any more than the inter-nuclear distances can do in
systems in which, using classical considerations, isomers are
possible. Likewise, the electronic energy of the problem as a
function of the translationally invariant nuclear variables can be
determined, as the expected value of the electronic
part of the full Hamiltonian (\ref{coulham}) obtained by
integrating over all the electronic space and spin coordinates.
But the electronic energy function will be invariant under any
permutation of like nuclei, so there will be no unique minimum in
it to be associated with an equilibrium geometry. The electronic
charge density and the electronic energy derived from the full
wave-function seem, therefore, to have properties quite different
from those that they have in the clamped-nuclei approximation. 
So, for the time being at least, it does not
seem that any charge density methods for identifying bonds,
can be regarded as properly based in the full problem.

It would seem that one cannot extract from the solutions of the full
problem many of those features that are desirable for chemical
explanations, but one can extract them from the clamped-nuclei
picture. However if it were possible to establish that the
clamped-nuclei Hamiltonian were an effective approximation to
$\mathsf{H}'$, that might be thought enough. 

\subsection{Clamping the nuclei}

It is sometimes asserted that the clamped-nuclei Hamiltonian can be
obtained from the Coulomb Hamiltonian by letting the nuclear masses
increase without limit. The Hamiltonian that would result if this were
done would be
\[\mathsf{H}^{\rm nn}({\mat x}^{\rm n} ,{\mat x}^{\rm e}) =
-\frac{\hbar^2}{2m}\sum_{i=1}^{N}
{\nabla}^2({\mat x}^{\rm e}_i) - \frac{e^2}{4\pi{\epsilon}_0}\sum_{i=1}^A
\sum_{j=1}^N\frac{Z_i}{|{\mat x}^{\rm e}_j - {\mat x}^{\rm n}_i|} + \frac{e^2}
{8\pi{\epsilon}_0}\sum_{i,j=1}^N\!\hbox{\raisebox{5pt}{${}^\prime$}}
\frac{1}{|{\mat x}^{\rm e}_i - {\mat x}^{\rm e}_j|}\]

\begin{equation}
+\frac{e^2}{8\pi{\epsilon}_0}\sum_{i,j=1}^A\!
\hbox{\raisebox{5pt}{${}^\prime$}} \frac{Z_iZ_j}{|{\mat x}^{\rm n}_i -
  {\mat x}^{\rm n}_j|}
\label{hnn}
\end{equation}
with formal Schr\"{o}dinger equation, by analogy with (\ref{cnp}),
\begin{equation}
\mathsf{H}^{\rm nn}({\mat x}^{\rm n} ,{\mat x}^{\rm e})\psi^{\rm
  nn}_p({\mat x}^{\rm n},{\mat x}^{\rm e})=E^{\rm nn}_p({\mat x}^{\rm
  n}) \psi^{\rm nn}_p({\mat x}^{\rm n} ,{\mat x}^{\rm e}).
\label{nnp}
\end{equation}

Given that the Coulomb Hamiltonian has eigenstates such that
\begin{equation}
 \mathsf{H}({\mat x}^{\rm n}, {\mat x}^{\rm e}){\psi}({\mat x}^{\rm
   n}, {\mat x}^{\rm e})
=E{\psi}({\mat x}^{\rm n}, {\mat x}^{\rm e})
\label{fp}
\end{equation}
then, if the solutions of (\ref{nnp}) were well defined, it would
seem that the eigenstates in (\ref{fp}) could be expanded as a sum of products of the form
\begin{equation}
\psi({\mat x}^{\rm n}, {\mat x}^{\rm e})= \sum_p{\Phi}_p({\mat x}^{\rm
  n}) {\psi}^{\rm nn}_p
({\mat x}^{\rm n}, {\mat x}^{\rm e})
\label{bh}
\end{equation}
where the \{$\Phi$\} play the role of `nuclear wavefunctions'.

In the Hamiltonian (\ref{hnn}) the nuclear variables are free and not
constant and there are no nuclear kinetic energy operators to dominate
the potential operators involving these free nuclear variables. The
Hamiltonian thus specified cannot be self-adjoint in the Kato
sense. It is certainly not the case either, that the nuclei variables
become constants, as asserted in Born and Oppenheimer. The Hamiltonian 
can be \textit{made} self-adjoint by clamping the
nuclei because the electronic kinetic energy operators can dominate
the potential operators which involve only electronic variables. The
Hamiltonian (\ref{hcn}) is thus a proper one. But since the Hamiltonian
(\ref{hnn}) is not self adjoint it is not at all clear that the
hoped for eigensolutions of (\ref{nnp}) form a complete set suitable for the expansion (\ref{bh}).
However that may be, it was observed more than 30 years ago and as we
have already seen here, that the arguments for
an expansion (\ref{bh}) are quite formal because the Coulomb
Hamiltonian has a completely continuous spectrum arising from the possibility of
uniform translational motion and so its solutions cannot be properly
approximated by a sum of this kind. This means too that the arguments of
Born and Oppenheimer and of Born  for his later
approach to representations of this kind, are also quite formal \cite{SW:12a,SW:12b}.

There is no need to specify the proposed $A-1$
translationally invariant nuclear variables $\mat{t}^{\rm n}$ other
than to say that they are expressed entirely in
terms of the laboratory nuclear coordinates. Of course the laboratory nuclear variable 
$\mat{x}^{\rm n}_i$ cannot be completely written in terms of the $A-1$ translationally invariant
coordinates arising from the nuclei, but in the electron-nucleus
attraction and in the nuclear repulsion terms the centre-of-nuclear
mass $\mat{X}$ cancels out. The  symbol $\mat{x}^{\rm n}_i$ will still be used to
denote the nuclear variables but it should be remembered that the
nuclear potentials are functions of the translationally invariant coordinates defined by the nuclear coordinates.

On  making this choice of electronic coordinates the Coulomb Hamiltonian
(\ref{coulham}) is transformed so that the the electronic part becomes:

\[\mathsf{H}^{' \rm e}(\mat x^{\rm n} ,{\mat t}^{\rm e}) =
-\frac{\hbar^2}{2m}\sum_{i=1}^{N}
{\nabla}^2({\mat t}^{\rm e}_i)
-\frac{\hbar^2}{2 M}\sum_{i, j=1}^{N}
\vec{\nabla}(\mat{t}^{\rm e}_i)\cdot\vec{\nabla}(\mat{t}^{\rm e}_j) -
\frac{e^2}{4\pi{\epsilon}_0}\sum_{i=1}^A
\sum_{j=1}^N\frac{Z_i}{|{\mat t}^{\rm e}_j - {\mat x}^{\rm n}_i|} \]
\begin{equation}
+ \frac{e^2}
{8\pi{\epsilon}_0}\sum_{i,j=1}^N\!\hbox{\raisebox{5pt}{${}^\prime$}}
\frac{1}{|{\mat t}^{\rm e}_i - {\mat t}^{\rm e}_j|} + \sum_{i,j=1}^A\!
\hbox{\raisebox{5pt}{${}^\prime$}} \frac{Z_iZ_j}{|\mat x^{\rm n}_i -
  \mat x^{\rm n}_j|}.
\label{hemol}
\end{equation}
This electronic Hamiltonian is translationally invariant and would yield the usual form were the
nuclear masses to increase without limit.

The nuclear part of the transformed Hamiltonian  involves only kinetic
energy operators and has the form:
\begin{equation}
\mathsf{K}^{\rm n}({\mat t}^{\rm n}) =
  -\frac{\hbar^2}{2}\sum_{i,j=1}^{A-1}\frac{1}{\mu^{\rm n}_
{ij}}\vec{\nabla}({\mat t}^{\rm n}_i).\vec{\nabla}({\mat t}^{\rm n}_j)
\label{hnt}
\end{equation}
with the inverse mass matrix suitably defined involving only the original nuclear variables.

Both (\ref{hemol}) and (\ref{hnt}) are invariant under any orthogonal
transformation of both the electronic and nuclear variables. If the
nuclei are clamped in (\ref{hemol}) then invariance remains only under
those orthogonal transformations of the electronic variables that
can be re-expressed as changes in the positions of nuclei with
identical charges while maintaining the same nuclear geometry.
The form (\ref{hemol}) remains invariant under all permutations
of the electronic variables and is invariant under permutation of the
variables of those nuclei with the same charge. Thus if an electronic
energy  minimum is found at some clamped nuclei geometry there will be
as many  minima as there are permutations of identically charged
nuclei. The kinetic energy operator (\ref{hnt}) is invariant under all
orthogonal transformations of the nuclear variables and under all
permutations of the variables of nuclei with the same mass.

The splitting of the translationally invariant Hamiltonian $\mathsf{H}'(\mat{t})$
into two parts breaks its symmetry, since each part exhibits only
a sub-symmetry of the full problem. If wavefunctions derived from
approximate solutions to (\ref{hemol}) are to be used to construct
solutions to the full problem (\ref{fp}) utilizing (\ref{hnt}) care will be needed
to couple the sub-symmetries to yield solutions with full
symmetry.

If the usual approach were taken to approximating solutions to the
nuclear motion Hamiltonian using sums of products of electronic and
nuclear parts a typical term in the sum used as trial function for
the form would be
\be \phi_{p}(\mat{t}^\re,
\mat{t}^\rn)\Phi_{p}(\mat{t}^\rn)
\label{spdtg}
\ee
where $p$ denotes an electronic state. The solutions are on the
Cartesian product space $R^{3A-3}\times R^{3N}$. There is no
explicit coupling of the nuclear motion and electronic motions and it is
thus possible to represent for any electronic
state, any number of rotational states. It is not generally possible to choose $\Phi$ directly as an
eigenfunction of the nuclear angular momentum, neither is it possible
to choose $\phi$ directly as an eigenfunction of the electronic
angular momentum. $\phi$ as usually computed belongs to the totally
symmetric representation of the symmetric group of each set of nuclei
with identical charges. $\Phi$ could then be a basis function for an
irrep of the symmetric group for each set of
particles with identical masses if the permutational symmetry were
properly considered in solving the nuclear motion problem.

Clamped-nuclei calculations are usually undertaken so as to yield a
potential that involves no redundant coordinates. Thus a
translationally invariant electronic Hamiltonian, as noted previously, 
would actually generate a more general potential than this.
A clamped-nuclei potential is therefore more properly associated with
the electronic Hamiltonian after the separation of rotational motion
than with the merely translationally invariant
one. With this choice again, 
\be \sum_{m=-J}^{J} {^J\phi}_{pm}(\mat{r}, \mat{R}
)^J\Phi_{pm}(\mat{R})|J M m>
\label{coupdtg}
\ee
where $\mat{R}$ represents the $3A-6$ internal coordinates invariant
under all orthogonal transformations of the $\mat{t}^\rn$ and
$|J M m>$ is an angular momentum eigenfunction. The general solutions
are on the manifold $R^{3A-6}\times S^3 \times R^{3N}$.

To achieve permutational symmetry in the nuclear motion part of the
wavefunction would in the general case be very difficult. The nuclei are
identified in the process of defining a body-fixed frame to
describe the rotational motion, even if they are identical. If only a
subset of a set of identical nuclei were used in such a definition,
some permutation of the nuclear variables would induce a change in the
definition of the body-fixed frame and thus spoil the rotational
separation. Thus permutations of identical nuclei are considered
usually only if such permutations correspond to point-group
operations which leave the body-fixing choices
invariant.

If one considers the clamped-nuclei Hamiltonian as providing input for
the full Hamiltonian in which the rotational motion is made explicit,
the basic nuclear motion problem should be treated as a $2J+1$
dimensional problem. If this is done then the translational and
rotational symmetries of the full problem are properly dealt
with. However the solutions are not generally basis functions for
irreps of the symmetric groups of sets of identical nuclei except for
such sub-groups as constitute the point groups used in frame
fixing. This restriction of the permutations is usually assumed to be
justified by appealing to the properties of the potential surface. The
idea here is widely believed and used in interpreting molecular spectra.

As noted earlier, the original attempts to justify  the
Born-Oppenheimer and the Born approaches from
the full Coulomb Hamiltonian, lack rigorous mathematical foundations.
So far there have been no attempts to make the foundations of the Born
approach mathematically secure. However the coherent states approach
has been used to give mathematically rigorous accounts of surface
crossings.  It seems very unlikely that it would be  possible to
provide a secure foundation for the Born approach in anything like the
manner in which it is usually presented.

The Born-Oppenheimer
approximation, whose validity depends on there being a deep enough
localized potential well in the electronic energy, has however been
extensively treated. The mathematical approaches depend upon the
theory of fibre bundles and the electronic Hamiltonian in these
approaches is defined in terms of a fibre bundle. It is central to
these approaches however that the fibre bundle should be trivial, that
is that the base manifold and the basis for the fibres be describable
as a direct product of Cartesian spaces. .

A mathematically satisfactory account of the Born-Oppenheimer
approximation for polyatomic in an approach based on (\ref{coupdtg})
has not yet been provided but it has proved possible to provide one
based on (\ref{spdtg}). Because the nuclear kinetic energy operator in
the space $R^{3A-3}$ cannot be expressed in terms of the nuclear
angular momentum, it is not possible in this formulation to separate
the rotational motion from the other internal motions. This work also
considers the possibility that there are two minima in the potential
as indeed there would be because of inversion symmetry if the
potential minimum were at other than a planar geometry. It does not,
however, consider the possibility of such multiple minima as might be
induced by permutational symmetry. It might be possible to extend the
two minima arguments to the multiple minima case and perhaps provide a
mathematically secure account of the potential energy surface
properties approach to ignoring some of the inconvenient permutations.
This has not so far been attempted.

At the same proper mathematical level it has been more recently
shown that almost any of the eigenvalues of the Coulomb Hamiltonian can be
approximated by eigenvalues of the clamped-nuclei Hamiltonian. The
correspondence does not depend on the clamped-nuclei eigenvalue being
one that corresponds to an electronic energy minimum.

There have also been claims that a bound-state wavefunction of the Coulomb
Hamiltonian can be written exactly as a single product rather than the sum
(\ref{bh}). But for exactly the same reasons that the initial
formulations of the Born-Oppenheimer and the Born approaches are purely
formal, so is this approach. It might be urged that this approach
would succeed if the translation motion were removed and the arguments
based on the use of $\mathsf{H}'(\mat{t})$ and the product (\ref{spdtg})
considered. But so far no mathematically sound formulation has been
made and there is a mathematically sound view that such a formulation
is very unlikely. 

For a secure account to be given in terms of the separation
(\ref{coupdtg}), which is what is really required if one is to use the
clamped-nuclei electronic Hamiltonian, it would be necessary to
consider more than one coordinate space. On the manifold $S^3$ as
explained before, once there are more than four particles, at least two coordinate spaces are required to span
the whole manifold because it is possible to construct two distinct
molecular geometries with the same internal coordinate specification
within a coordinate space, so
that a potential expressed in the internal coordinates cannot be
analytic everywhere. It would therefore seem to be an impossible 
job. But even if it were to be accomplished it seems very unlikely
that a multiple minima argument could be constructed to account for
point-group symmetry in this context. It is possible to show, that in
the usual form of the Hamiltonian for nuclear motion, where the axes
are defined in terms of a given nuclear geometry,
permutations can be such as to cause the body-fixed frame definition
to fail completely.

Naturally any extension of the trial wave function for the full Coulomb Hamiltonian
problem from a single term to a many term form must be welcomed as an
advance; it is however simply a technical advance and it might prove premature
to load that technical advance with too much physical import.

At present it is not possible properly to place the clamped-nuclei electronic
Hamiltonian in the context of the full problem, including nuclear
motion. However if the nuclei were treated as distinguishable
particles, even when formally identical, then some of difficulties that
arise from the consideration of nuclear permutations would not
occur. But it would still be necessary to be able to justify the
choice of sub-sets of permutations among identical particles when such
seem to be required to explain experimental results. A particular
difficulty arises here for it is not possible to distinguish between
isomers nor is it possible to specify a molecular geometry, unless 
it is possible to distinguish between formally identical particles.

But regarding the nuclei as distinguishable would not avoid the difficulty of
constructing total angular momentum eigenfunctions from the nuclear
and electronic parts. Such treatment of the nuclei would not make the
traditional demonstrations of the Born-Oppenheimer or the Born
approximations mathematically sound either. However it would ensure
that the mathematically sound presentations of the Born-Oppenheimer
approximation mentioned earlier need no further extension to include
permutations of identical nuclei. There is, unfortunately, little good
to be said, from a mathematical point of view, of the traditional Born
argument. This is troubling because the Born approach is assumed to
provide the basis for the consideration of chemical reactions on and
between potential energy surfaces. However it is clear that the
clamped-nuclei (electronic) Hamiltonian can be usefully deployed in
nuclear motion calculations if the nuclei are considered identifiable.

\section{Discussion}
We have suggested in the foregoing that using the clamped-nuclei
Hamiltonian, treating the nuclei as classical
particles while calculating electronic wavefunctions, and then using
the electronic functions as a basis for a semiclassical treatment of
nuclear motion, will lead to a coherent picture that is compatible with
classical chemical explanation.

At the start of the 21$^{\mbox{st}}$ century Simon presented a list of open problems in
mathematical physics among which Problem 12 is of relevance here \cite{BS:00}:
\begin{quote}
{\bf Problem 12}: Is there a mathematical sense in which one can justify from first principles current
techniques for determining molecular configurations ?
\end{quote}
This problem, although stated in mathematically vague terms,
should be viewed as asking for some precise way to go from
fundamental quantum theory to configurations of molecules;
evidently Simon did not see \textit{ab initio} electronic
structure theory as a complete answer. We do not know exactly what Simon envisaged 
with ``from first principles'', but it is plausible he had the Coulomb Hamiltonian in mind.
In our view, and contrary to L\"{o}wdin's expectation, \S \ref{msiso}, we see no reason to 
suppose that chemistry can be founded on the Coulomb Hamiltonian. Let us first summarize
briefly what we see as the outstanding difficulties.

We should note at the outset that in our account of quantum chemistry, we have not mentioned 
any aspects of special relativity except, incidentally, spin. Spin provides the basis
for some spectroscopic methods of chemical importance such
as nuclear magnetic resonance, but such behaviour can be treated
perfectly adequately by first-order perturbation theory so our
discussion can be regarded as complete. Other relativistic
considerations seem irrelevant to chemical explanation. Indeed we have no idea of how a quantum 
mechanical account of a defined collection of electrons and nuclei - as required
for the minimal specification of a molecule - with electromagnetic interactions, that is
Lorentz invariant, can be developed.

The most that can be shown is that, over a limited energy range, certain of
the bound-state energies of the Coulomb Hamiltonian can be
well-approximated by eigenvalues of the clamped-nuclei Hamiltonian; however the conventional
product of the electronic wavefunctions (from the clamped-nuclei Hamiltonian) and
associated nuclear wavefunctions lack the symmetry properties of Coulomb
Hamiltonian eigenfunctions and this difference has not been explained.
It will be a matter of some mathematical difficulty to show that the
full allowed nuclear permutation group has irreps which can be
approximated by the irreps of the permutational sub-group which 
characterises a given nuclear geometry. It will be a project of
similar difficulty to show that the full rotation-inversion group has
irreps that are approximated by the results obtained from the combined nuclear
and electronic treatment arising from the clamped-nuclei
Hamiltonian. Such results will have rotation-inversion symmetry only by
accident and so irreps are not usually definable. In neither case is it
likely that a general demonstration will be possible and the
expectation is that any results must be achieved on a case by case
basis. It is quite clear that the eigenvalues/eigenfunctions of the Coulomb 
Hamiltonian itself do not lend themselves to classical chemical explanations. So the best that
can be hoped for in the relationship between chemical explanation and
explanation in terms of results from the Coulomb Hamiltonian, is a rather tenuous one.

In some contexts the quantum mechanical properties of nuclei are crucial; in other situations 
they are simply dropped, and nuclei are treated as classical, distinguishable particles. Thus, for example,
the fact that the deuteron, D, and the N$^{14}$ nucleus are \textit{boson} particles can be recognized
from characteristic features in the band spectra of D$_{2}$ and N$_{2}$ respectively. The spectra of
diatomic molecules are traditionally described using the clamped-nuclei approach (\S \ref{compQC}) in terms
of potential energy curves and the spin of the nuclei has to be added \textit{ad hoc}; however it is 
perfectly possible to use a moving-nuclei description (\S \ref{tfp}) 
of diatomic molecules in which the boson/fermion classification of nuclei sits comfortably.

In most of chemistry however there is no reference to this basic aspect of quantum mechanics. 
As an example\footnote{This is \textit{not} an isolated special case; it is generic.} there are 
three familiar isomers of formula 
C$_{3}$H$_{4}$; the structural principle predicts correctly that replacement of two hydrogen atoms by two
deuterium atoms leads to seven distinct species, two of which are predicted to be optically active. All
 seven have been synthesized in the laboratory, including the resolved enantiomers. About that 
 quantum mechanics based on the Coulomb Hamiltonian (\S \ref{ClHam}) apparently has nothing to say.

This suggests that the Coulomb Hamiltonian on its own (the `isolated molecule model') is
not an adequate basis for a quantum mechanical account of chemistry, and so one is led to
consider the role of persistent interactions of an environment with the charged particles
constituting a molecule. Various suggestions as to how the `environment' is to be
characterized can be found in the literature, for example: other molecules, `thermal baths', the
quantized electromagnetic field \textit{etc.}, together with finite temperatures; such 
discussions largely focus on models far removed from the Coulomb 
Hamiltonian\cite{SW:05,SW:12,SW:03,EBD:95}. None of them really explain, in quantum 
mechanical terms, how one gets from indistinguishable identical particles to 
(classical) distinguishable particles, other than by putting the requisite answer in by hand.

Our account, which we believe is an accurate and impartial one, is
deeply puzzling. One could of course elucidate matters by saying that there
must be another theory. But that wouldn't help at the moment and it
seems a sensible use of epistemological imagination on the puzzle
would be very welcome. But that is a matter in which we believe
philosophers of science are much more capable that are we.

\end{document}